\documentclass[aps,prb,twocolumn,amsmath,amssymb]{revtex4}
\usepackage{graphicx} 
\usepackage{dcolumn} 
\usepackage{bm} 
\usepackage{epsfig}
\usepackage{amsmath}
\usepackage{amssymb}
\usepackage{hyperref}
\begin{document}
\newcommand{\bsy}[1]{\mbox{${\boldsymbol #1}$}} 
\newcommand{\bvecsy}[1]{\mbox{$\vec{\boldsymbol #1}$}} 
\newcommand{\bvec}[1]{\mbox{$\vec{\mathbf #1}$}} 
\newcommand{\btensorsy}[1]{\mbox{$\tensor{\boldsymbol #1}$}} 
\newcommand{\btensor}[1]{\mbox{$\tensor{\mathbf #1}$}} 
\newcommand{\tensorId}{\mbox{$\tensor{\mathbb{\mathbf I}}$}} 
\newcommand{\be}{\begin{equation}}
\newcommand{\ee}{\end{equation}}
\newcommand{\bea}{\begin{eqnarray}}
\newcommand{\eea}{\end{eqnarray}}
\newcommand{\e}{\mathrm{e}}
\newcommand{\arccot}{\mathrm{arccot}}
\newcommand{\arctanh}{\mathrm{arctanh}}
\title{Brewster solitons and omnidirectional solitons in one-dimensional optical structures involving Kerr-type nonlinear materials.}
\author{J. Posada-Loaiza,$^{1}$ J. D. Mazo-V\'asquez,$^{1,2,3}$ and E. Reyes-G\'omez$^{1}$}
\email{$^{*}$amador.reyes@udea.edu.co} 
\address{$^{1}$Instituto de F\'{i}sica, Universidad de Antioquia - UdeA, Calle 70 No. 52-21, Medell\'{\i}n, Colombia,}
\address{$^{2}$Max Planck Institute for the Science of Light, 91058 Erlangen, Germany,}
\address{$^{3}$Department of Physics, Friedrich-Alexander-Universit\"at Erlangen-N\"urnberg, 91058 Erlangen, Germany.}
\begin{abstract}
The possibility of exciting stationary solitons due to the Brewster effect is investigated. The analysis is performed by computing the transmission coefficient of an electromagnetic monochromatic wave obliquely incident on the optical system and linearly polarized with the transverse-electric polarization. The optical system is supposed to be surrounded by a vacuum. Two different optical systems are studied. First, the system under consideration is assumed to be composed of the junction of a single dispersive layer with a linear electromagnetic response and a single nonlinear layer exhibiting Kerr nonlinearity. The obtained results suggest the excitation of stationary solitons in the nonlinear layer induced by the occurrence of the Brewster phenomenon in the linear medium. The second system is considered as a single Kerr nonlinear layer. The Brewster effect is observed for both self-focusing and self-defocusing nonlinearities,  which leads to the excitation of soliton states within the nonlinear layer. The  Brewster-angle dependence of the intensity of the incident electromagnetic wave is also shown. Finally, the excitation of omnidirectional solitons in the case of single nonlinear layers is discussed. It is shown that matching the magnetic permeability of the surrounding medium and the nonlinear layer leads to maximum transmission states, independent of the incidence angle, for the input-intensity value that causes the homogenization of the electric permittivity.
\end{abstract}
\date{\today}
\maketitle
\section{Introduction}
\label{intro}

In 1815, Sir David Brewster realized that when an unpolarized electromagnetic wave is incident on the interface between two optical media, for a specific value of the incidence angle, the reflected wave is polarized in the direction perpendicular to the plane of incidence \cite{Brewster1815}. Nowadays, such a phenomenon is known as the Brewster effect. If the incident electromagnetic wave is linearly polarized, then the Brewster effect may lead, under some circumstances, to the suppression of the reflected wave. For instance, the Brewster phenomenon occurring at the interface between two non-magnetic natural dielectrics leads to the suppression of the reflected wave when the incident wave is polarized in the transverse-magnetic (TM) configuration, i.e., for $p-$polarized incident waves. In this scenario, if the absorption coefficient of the refracting medium is negligible, the energy conservation law ensures that the wave-transmission coefficient through the interface reaches maximum values. A similar situation may be observed for incident waves in the transverse-electric (TE) configuration ($s-$polarization). In this case, at least one of the materials that make up the interface must exhibit a strong dispersion and magnetic activity at the frequency of the incident wave \cite{TamayamaPRB2006}. Adding optical-active two-dimensional layers at the interface between two non-magnetic dielectrics may also lead the Brewster effect to occur \cite{LinPRA2016,SreekanthACSP2019}.

The study of the Brewster phenomenon also has practical applications in optical engineering and polarization technologies \cite{JungkPM1994}. Most are based on exploiting maximum transmission states in the optical system. For instance, in the case of disordered optical systems, even though disorder should lead to the existence of localized states within the optical system, the Brewster effect allows the propagation of delocalized electromagnetic waves known as Brewster anomalies \cite{AronovSSC1990,AsatryanPRL2007,MogilevtsevPRB2010}. Similarly, the Brewster effect could generate maximum transmission states through linear/nonlinear optical systems. Suppose, for example, an electromagnetic wave incident from the vacuum on a linear material (refracting medium). If a nonlinear optical medium is added to the refracting medium, the possibility of obtaining high-transmission refracted electromagnetic waves via the Brewster effect could be exploited to generate soliton waves in the nonlinear medium. Due to the suppression of the reflected wave, one could expect maximum-transmission states in the nonlinear medium. It could be interesting to examine the case where the incidence occurs directly on the nonlinear layer. Such a case is formally equivalent to suppressing the dispersive linear medium, and the occurrence of the Brewster effect on the vacuum/nonlinear medium interface could straightforwardly lead to the excitation of soliton states within the nonlinear optical system.

The present paper aims to investigate, in linear/nonlinear optical systems, the possibility of combining the occurrence of the Brewster effect in the linear medium and the consequent generation of stationary solitons in the nonlinear optical layers. Moreover, we study the Brewster effect in the case of a single nonlinear layer and the resultant formation of soliton states. This paper is organized as follows: In section \ref{theory}, we describe the theoretical procedure used to describe the light propagation throughout the optical linear/nonlinear system. Numerical results are described in section \ref{results}. Finally, our conclusions are given in section \ref{conclusions}.

\section{Theoretical framework}
\label{theory}

\begin{figure} [t]
\centering
\begin{tabular}{rl}
\includegraphics[width=0.5\columnwidth]{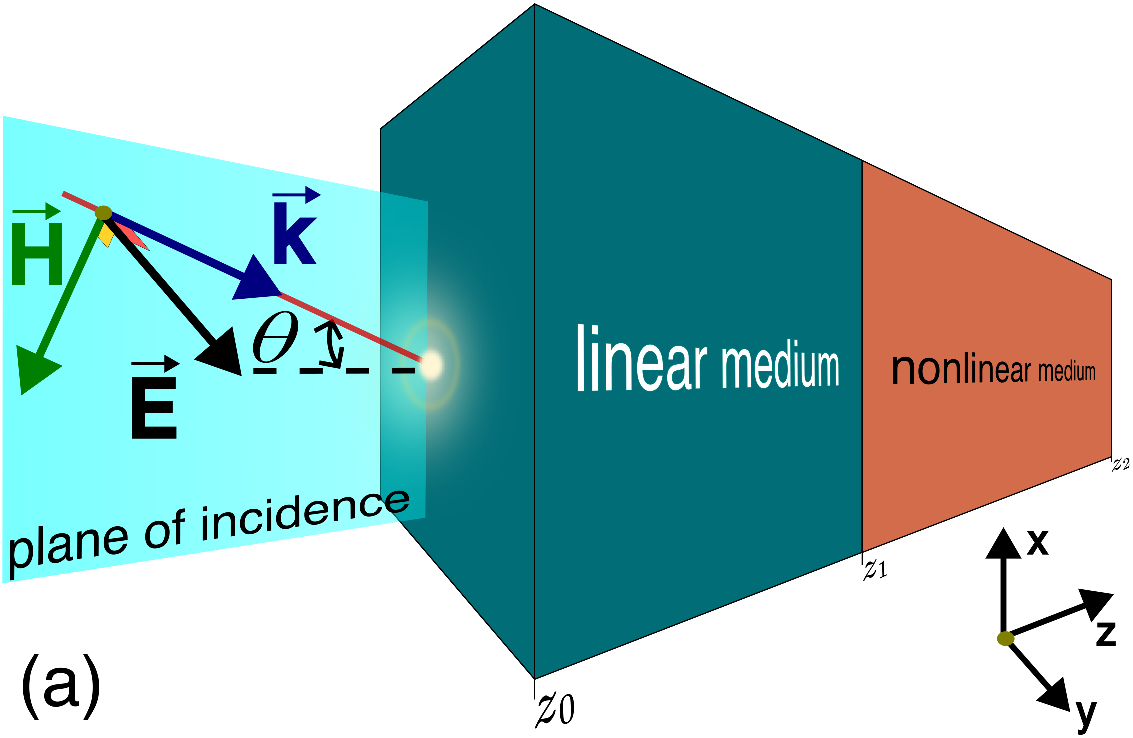} &
\includegraphics[width=0.47\columnwidth]{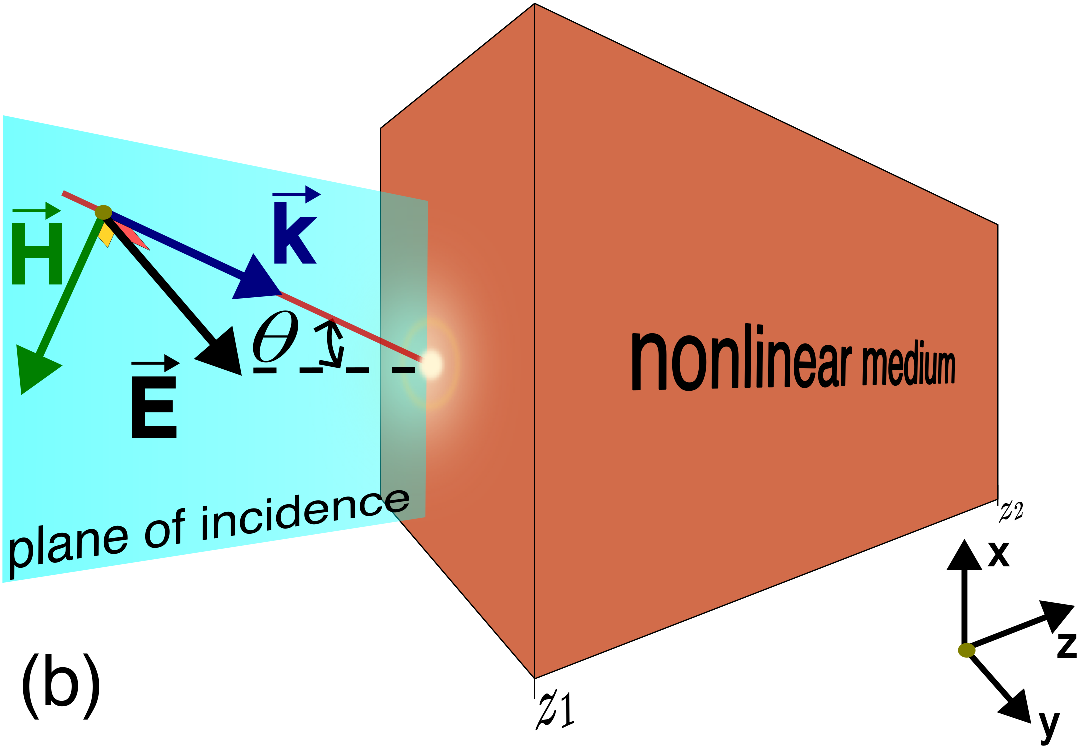}
\end{tabular}
\caption{(Color online)  A pictorial view of the optical systems is under consideration. The electromagnetic radiation of frequency $\omega$ is obliquely incident upon the system in the transverse-electric configuration. Panel (a) displays the case of a linear/nonlinear bilayer system as described in the text. Panel (b) corresponds to the incidence upon a single nonlinear layer.}
\label{fig1}
\end{figure}

We consider first an electromagnetic wave, with frequency $\omega$, incident upon an optical system in the TE configuration, as displayed in Fig. \ref{fig1}(a). We assume the system comprises a stack of materials grown along the $z$ direction. The stack includes an optical material with linear electromagnetic responses and a single layer of a nonlinear-Kerr material. The incidence is supposed to take place on the linear layer, which displays magnetic activity to allow the occurrence of the Brewster effect for linearly polarized incident waves with TE polarization. The whole system is surrounded by a vacuum, with electric permittivity $\varepsilon_0 = 1$ and magnetic permeability $\mu_0 = 1$.

\begin{figure}
\centering
\includegraphics[width=0.5\columnwidth]{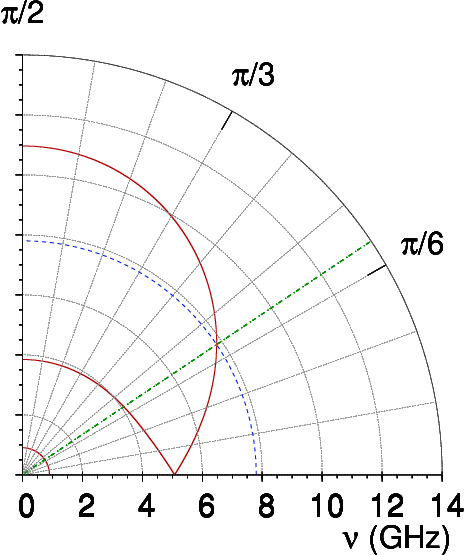} 
\caption{(Color online)  The Brewster line $\nu = \nu(\theta_B)$ obtained from the solution of the transcendental Eq. \eqref{eq5}, corresponding to the electromagnetic responses given by Eqs. \eqref{eq11} and \eqref{eq12}.  The dashed line at the frequency $\nu = 7.8$  GHz corresponds to the Brewster angle $\theta_{B} = 33.82^{\circ}$ (0.59 rad), which is represented as a dotted-dashed oblique line.}
\label{fig2}
\end{figure}

The electric field of the electromagnetic wave in the TE configuration may be written as $\bvec{E}(\bvec{r},t)=E(z) \exp \left [ i (q x - \omega t) \right ] \bvec{e}_y$, where $q = \frac{\omega}{c} \sqrt{\varepsilon_0 \mu_0}\sin (\theta)$ is the component of the wave vector parallel to the interface along the $x$ direction, $\theta$ the incidence angle relative to the vacuum, and $\bvec{e}_y$ is the unit vector along the $y$ direction. According to the Maxwell equations, the amplitude of the electric field $E = E(z)$ satisfies the differential equation \cite{CavalcantiOL2014}

\begin{equation}
\label{eq1}
    -\frac{d }{dz}\left[ \frac{1}{\mu(z)}\frac{d }{dz} E(z) \right] - \left[ \frac{\omega^2}{c^2}\varepsilon(z) - \frac{q^2}{\mu(z)} \right]E(z) = 0,
\end{equation}
where $\varepsilon =  \varepsilon(z)$ and $\mu = \mu(z)$ are the whole system growth-direction-dependent electric permittivity and magnetic permeability, respectively.

Within the linear layer, one has $\mu (z) = \mu_{l}$ and $\varepsilon (z) = \varepsilon_{l}$, where both the electric permittivity $\varepsilon_{l}$ and magnetic permeability $\mu_{l}$ are considered to be functions of the wave frequency as detailed below, but independent on the position. Within the nonlinear slab, we take the magnetic permeability $\mu (z) = \mu_{nl}$ as independent of both the position and the wave frequency, and the electric permittivity given by
\be
\label{eq2}
\varepsilon (z) = \varepsilon_{nl} (z) = \varepsilon_{nl}^0 + \alpha \vert E(z)\vert^2,
\ee
where $\varepsilon_{nl}^0 $ is the background electric permittivity and $\alpha$ is the nonlinear dielectric coefficient. Eq. \eqref{eq2} results from the third-order expansion of the nonlinear polarization in a series of powers of the electric field by considering the nonlinear material as isotropic \cite{Sutherland2003,Kivshar2003}. The cases $\alpha>0$ and $\alpha<0$ correspond to self-focusing and self-defocusing nonlinear materials, respectively.

Eq. \eqref{eq1} may be solved by proposing the electric-field amplitude as
\begin{widetext}
\be
\label{eq3}
E(z) = E_i 
\begin{cases}
    \exp \left [ i  Q_0  \left ( z-z_0 \right ) \right ] + r  \exp \left [ -i  Q_0  \left ( z-z_0 \right ) \right ]  & \text{if } z < z_{0}, \\
    \psi(z) & \text{if } z_0 \leq z < z_1 , \\
    \phi(z) & \text{if } z_1 \leq z < z_2 , \\
    t \exp \left [ i  Q_0  \left ( z-z_2 \right ) \right ]  & \text{if } z > z_2,
\end{cases}
\ee
\end{widetext}
where $Q_0 = \frac{\omega}{c} \sqrt{\varepsilon_0 \mu_0} \cos (\theta)$, $E_i$ is the electric-field amplitude of the incident wave, $r$ and $t$ are the amplitudes of the reflected and transmitted wave, respectively, whereas $\psi$ and $\phi$ are the position-dependent amplitudes of the electric field inside the linear and nonlinear layers, respectively, in units of $E_i$. Notice (cf. Fig. \ref{fig1}(a)) that the surrounding medium (the vacuum) corresponds to the regions $z < z_{0}$ and $z > z_2$, whereas $z_0 \leq z < z_1$ ($z_1 \leq z < z_2$) corresponds to the linear (nonlinear) layer. We adopt the transfer-matrix formalism within the linear layer according to Ref. \onlinecite{CavalcantiPRB2006}. Within the nonlinear layer Eq. \eqref{eq1} becomes \cite{HincapieEPL2021}
\bea
\label{eq4}
\frac{d^2}{dz^2} \phi(z) + \frac{\omega^2}{c^2} \left [ \varepsilon_{nl}^0 \mu_{nl} - \varepsilon_0 \mu_0 \sin^2 (\theta) \right ] \phi(z) \nonumber \\ + \frac{\omega^2}{c^2} \mu_{nl} a \vert \phi(z) \vert^2 \phi(z) = 0, 
\eea
where $a = \alpha \vert E_i \vert^2$ is the normalized input intensity \cite{HincapieEPL2021}. Eq. \eqref{eq4} may be solved analytically in terms of the Jacobi elliptic functions \cite{HincapieEPL2021} or numerically by using the fourth-order Runge-Kutta method \cite{CavalcantiOL2014,YepesEPJD2024}, among other procedures.

We impose the continuity of both $E(z)$ and $\frac{1}{\mu(z)} \frac{d}{dz} E(z)$ at all the interfaces $z_0$, $z_1$, and $z_2$. As a result, the reflection and transmission coefficients may be obtained from the expressions $R = r r^*$ and $T = tt^*$, respectively. A maximum value of the transmission coefficient $T$ is expected to occur under the fulfillment of the Brewster condition corresponding to the TE polarization, i.e., 
\be
\label{eq5}
\sin^2 (\theta_B) = \frac{\frac{\varepsilon_l (\nu)}{\varepsilon_0}-\frac{\mu_l (\nu)}{\mu_0}}{\frac{\mu_0}{\mu_l (\nu)}-\frac{\mu_l (\nu)}{\mu_0}},
\ee
where $\theta_B \neq 0$ is the so-called Brewster angle and $\nu = \frac{\omega}{2 \pi}$ is the linear frequency of the incident wave. The Brewster effect is expected to occur at the interface $z_0$  under the condition that $\mu_l (\nu) \neq \mu_0$ \cite{LinPRA2016,SreekanthACSP2019}. Eq. \eqref{eq5} implicitly defines the relation $\nu = \nu(\theta_B)$ (the Brewster line), depending on the properties of the linear medium and allowing one to anticipate the expected Brewster angle for a given value of the incident-wave frequency. 

\begin{figure}
\centering
\begin{tabular}{rl}
\includegraphics[width=0.5\linewidth]{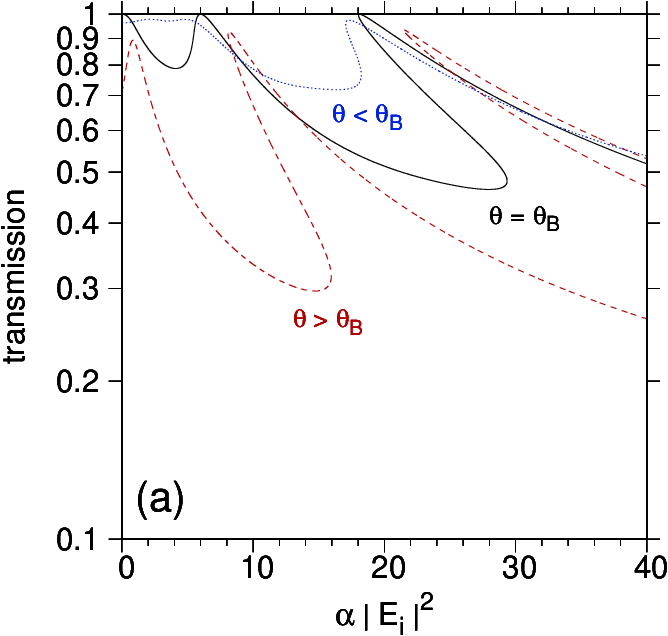} &
\includegraphics[width=0.5\linewidth]{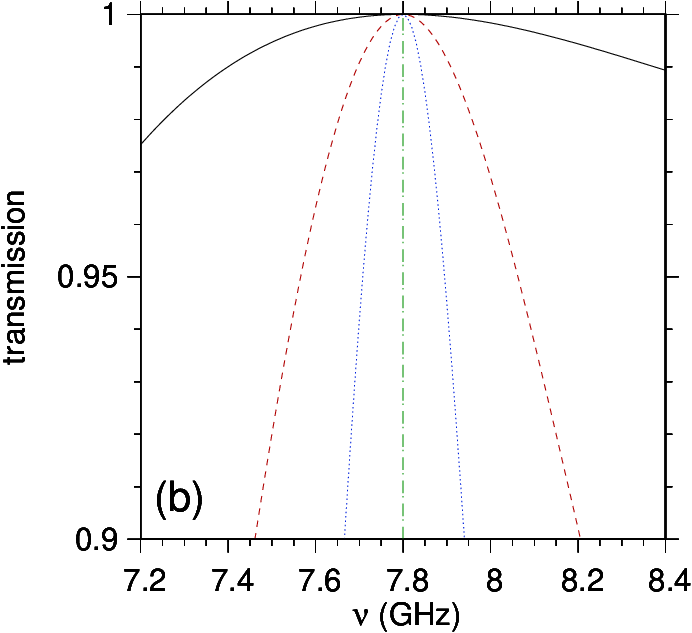}
\end{tabular}
\caption{(Color online)  Transmission coefficient as a function of (a) the normalized input intensity and (b) the incident-wave frequency for a linear/nonlinear double-layer system of widths $d_l = d_{nl} = 10$ mm, in the case focusing nonlinearity $\alpha > 0$. Results of panel (a) were obtained for $\nu = 7.8$  GHz corresponding to the Brewster angle $\theta_{B} = 33.82^{\circ}$ (0.59 rad). Solid, dashed, and dotted lines in panel (a) correspond to the incidence angles $\theta = \theta_{B}$, $\theta = 53.82^{\circ}$, and $\theta = 13.82^{\circ}$, respectively. The panel (b) results were obtained for $\nu = 7.8$ GHz and the corresponding Brewster angle $\theta_{B}$. Solid, dashed and dotted lines in panel (b) were obtained for $\alpha \vert E_{i} \vert^2$  = 0.00, 6.01, and 18.12, respectively, corresponding to the three first local maxima of the transmission coefficient displayed in panel (a), for  $\nu = 7.8$  GHz and $\theta_{B} = 33.82^{\circ}$ (cf. solid line in panel (a)). The vertical dotted-dashed line in panel (b) corresponds to  $\nu = 7.8$  GHz.}
\label{fig3}
\end{figure}

A particular case of interest that results from removing the linear material layer in the optical system may also be considered. It means the surrounding vacuum replaces the linear medium, so the incidence interface between the optical system and the vacuum is now at the position $z_1$ (see Fig. \ref{fig1}(b)). In this case, the electric-field amplitude in Eq. \eqref{eq3} can be rewritten as
\begin{widetext}
\be
\label{eq6}
E(z) = E_i 
\begin{cases}
    \exp \left [ i  Q_0  \left ( z-z_0 \right ) \right ] + r  \exp \left [ -i  Q_0  \left ( z-z_0 \right ) \right ]  & \text{if } z < z_{1}, \\
    \phi(z) & \text{if } z_1 \leq z < z_2 , \\
    t \exp \left [ i  Q_0  \left ( z-z_2 \right ) \right ]  & \text{if } z > z_2.
\end{cases}
\ee
\end{widetext}

The position-dependent electric-field amplitude within the nonlinear layer depends on the normalized input intensity through Eq. \eqref{eq4}. The corresponding analytical solution may be proposed as $\phi (z) = f (z) \exp \left [i \varphi (z) \right ]$, where $f$ and $\varphi$ are real functions of $z$. The square $F(z) = f^2 (z)$ of the real part of $\phi$ satisfies \cite{HincapieEPL2021}
\be
\label{eq7}
\lim_{a \rightarrow \frac{a_0}{F(z_2)}} F(z) = F(z_2), 
\ee
where
\be
\label{eq8}
a_0 = \varepsilon_0 \frac{\mu_{nl}}{\mu_0} \cos^2 (\theta) + \varepsilon_0 \frac{\mu_{0}}{\mu_{nl}} \sin^2 (\theta) - \varepsilon_{nl}^0.
\ee
Moreover, $F(z_2)$ is the square of the electric-field amplitude of the wave at the outgoing interface. The above equation was carefully demonstrated in Ref. \onlinecite{HincapieEPL2021}. We refer the reader to that reference and its corresponding supplementary material for mathematical details. According to such a procedure, the transmission coefficient is precisely $T = F(z_2)$. Consequently, the condition of maximum transmission ($T=1$) is reached at the limit $a \rightarrow a_0$ or, equivalently,
\be
\label{eq9}
\varepsilon_0 \frac{\mu_{nl}}{\mu_0} \cos^2 (\theta) + \varepsilon_0 \frac{\mu_{0}}{\mu_{nl}} \sin^2 (\theta) - \varepsilon_{nl}^0 = a = \alpha \vert E_i \vert^2.
\ee
Notice that, as $T + R = 1$, the condition $T=1$ implies $R=0$ and, consequently, $r = 0$ in Eq. \eqref{eq6}. Then, one has $E(z_1) = E_i$. Moreover, the incidence angle $\theta \neq 0$ may be interpreted as the Brewster angle $\theta_B$. Therefore, Eq. \eqref{eq9} may be rewritten as
\be
\label{eq10}
\sin^2 (\theta_B) = \frac{\frac{\varepsilon_{nl} (z_1)}{\varepsilon_0}-\frac{\mu_{nl}}{\mu_0}}{\frac{\mu_0}{\mu_{nl}}-\frac{\mu_{nl}}{\mu_0}},
\ee
where $\varepsilon_{nl} (z_1) = \varepsilon_{nl}^0 + \alpha \vert E_i\vert^2$. The above equation is similar to Eq. \eqref{eq5} but describes the Brewster effect at the interface $z_1$ between vacuum and the nonlinear medium for TE-polarized electromagnetic waves, obliquely incident upon the system. This equation corrects a misprint in the Eq. (7) of Ref. \onlinecite{YepesEPJD2024}.

\section{Results and discussion}
\label{results}

\begin{figure}
\centering
\begin{tabular}{rcl}
\includegraphics[width=0.33\linewidth]{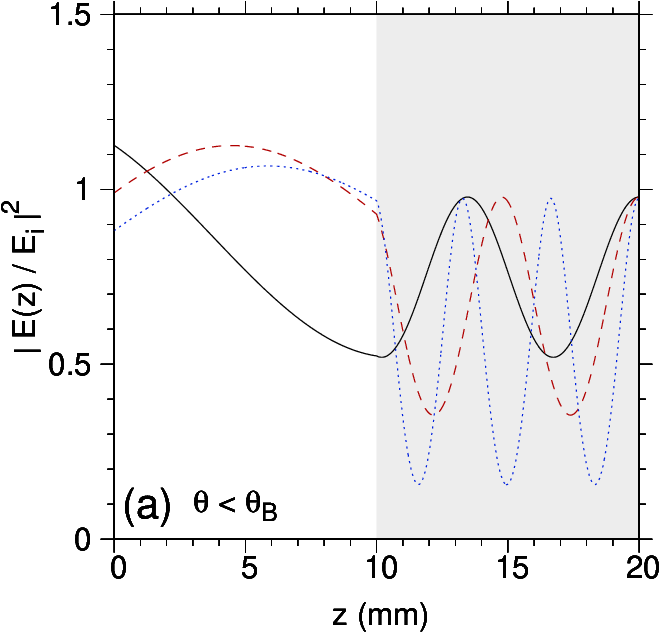} & 
\includegraphics[width=0.33\linewidth]{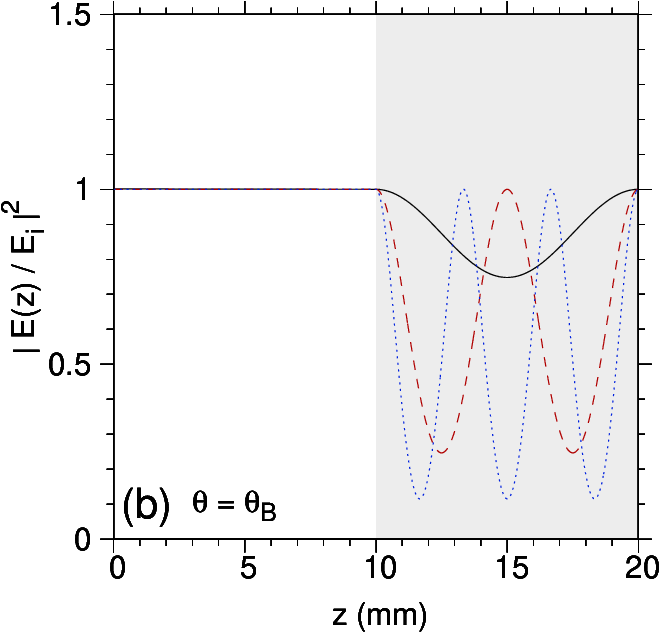} &
\includegraphics[width=0.33\linewidth]{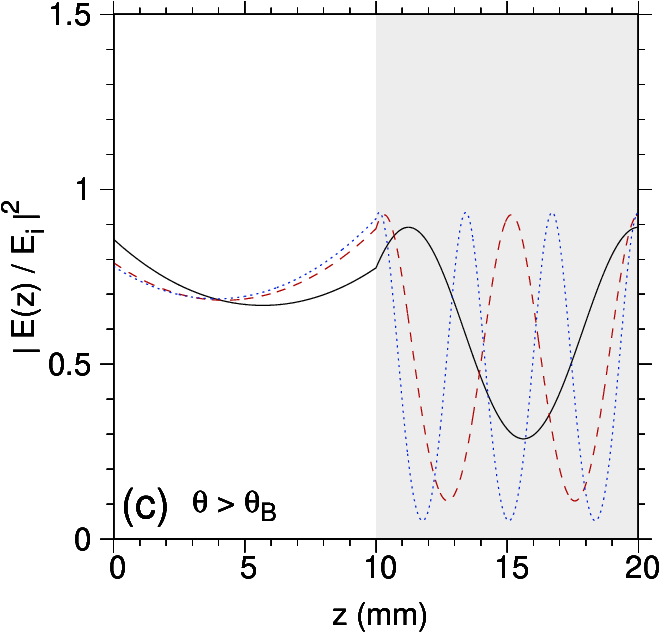}
\end{tabular}
\caption{(Color online)  Square of the electric-field amplitude as a function of the position within the optical system for $\nu = 7.8$ GHz and (a) $\theta = 13.82^{\circ}$, (b)  $\theta = \theta_{B}=33.82^{\circ}$ corresponding to the Brewster angle, and (c) $\theta = 53.82^{\circ}$. In each panel, solid, dotted, and dashed lines were computed for the values of $\alpha |E_{i}|^2$ corresponding to the three first peaks of the transmission coefficient displayed in Fig. \ref{fig3}(a) for the respective value of the incidence angle  ($\alpha |E_{i}|^2$  = 2.10, 4.84, and 17.31, respectively, in panel (a), $\alpha |E_{i}|^2$  = 0.0, 6.01, and 18.12, respectively, in panel (b), and $\alpha |E_{i}|^2$  = 0.83, 8.15, and 21.48, respectively, in panel (c)).} 
\label{fig4}
\end{figure}

\begin{figure}
\centering
\begin{tabular}{rl}
\includegraphics[width=0.5\linewidth]{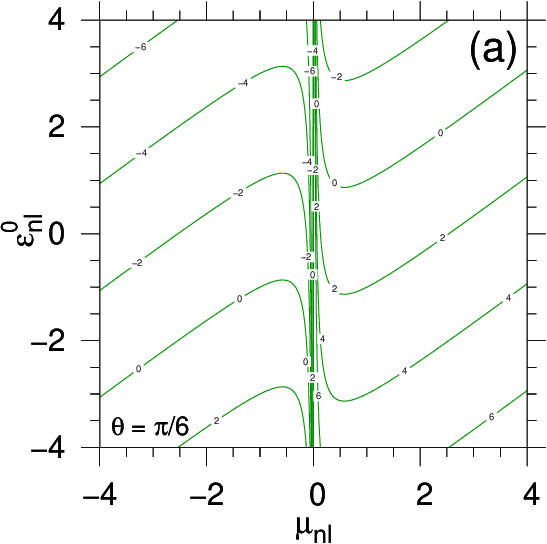} &
\includegraphics[width=0.5\linewidth]{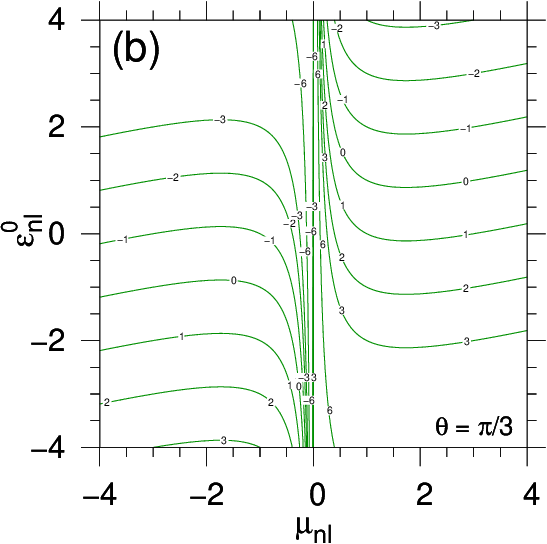}
\end{tabular}
\caption{(Color online)  Background electric permittivity $\varepsilon_{nl}^0$ as a function of magnetic permeability $\mu_{nl}$ (cf. Eq. \eqref{eq9}) guarantying the fulfillment of Eq. \eqref{eq10} for (a) $\theta = \pi/6$ and (b) $\theta = \pi/3$. Different curves represent the contour lines of the normalized input intensity $a = \alpha \vert E_i \vert^2$. Results were obtained for $\varepsilon_0 = \mu_0 = 1$ (the surrounding medium is a vacuum), for both self-focusing ($a>0$) and self-defocusing ($a<0$) nonlinear materials as well as for the linear regime ($a=0$).}
\label{fig5}
\end{figure}

\subsection{Linear/nonlinear optical structure}

First, we consider the linear/nonlinear optical system composed of a single dispersive layer with a linear electromagnetic response and a single nonlinear layer exhibiting Kerr nonlinearity. The frequency-dependent electric permittivity and magnetic permeability of the linear material are given by the expressions \cite{HegdeOL2005,ReyesSM2016}
\be
\label{eq11}
\varepsilon_l (\nu) = f_{\varepsilon} + \frac{F_{\varepsilon}}{\beta_{\varepsilon} - \nu^2}
\ee
and
\be
\label{eq12}
\mu_l (\nu) = f_{\mu} + \frac{F_{\mu}}{\beta_{\mu} - \nu^2},
\ee
respectively, where $\nu$ is the wave frequency in GHz, $f_{\varepsilon}  = 1.6$, $F_{\varepsilon} = 40 $ GHz$^2$, $\beta_{\varepsilon}=0.81$ GHz$^2$, $f_{\mu} = 1.0$, $F_{\mu} = 25$ GHz$^2$, and $\beta_{\mu} = 0.814$ GHz$^2$. Within the nonlinear layer we take \cite{ReyesSM2016} $\mu_{nl} = 2$ and $\varepsilon^0_{nl} = 2$ in Eq. \eqref{eq2}. For simplicity, no absorption effects in the optical system are considered. The thicknesses of the linear and nonlinear layers are $d_l = z_1 - z_0 = 10$ mm and $d_{nl} = z_2 - z_1 = 10$ mm, respectively.

We show in Fig. \ref{fig2} the relation $\nu = \nu(\theta_B)$, obtained from the solution of the transcendental Eq. \eqref{eq5} by taking into account that Eqs. \eqref{eq11} and \eqref{eq12} give the electromagnetic response of the linear dispersive medium.  The relation $\nu = \nu(\theta_B)$ indicates the frequency of the TE-polarized electromagnetic wave necessary for the Brewster phenomenon to occur at the incidence angle $\theta = \theta_B$. For numerical purposes, we have chosen $\nu = 7.8$  GHz corresponding to the Brewster angle $\theta_{B} = 33.82^{\circ}$ (0.59 rad approximately).

We display in Fig. \ref{fig3} the transmission coefficient as a function of both the normalized input intensity and wave frequency (cf. Figs. \ref{fig3}(a) and \ref{fig3}(b), respectively). In the aim of simplicity, calculations were performed by assuming the nonlinear layer made of a self-focusing material ($\alpha > 0$). The linear regime $\alpha=0$ is also computed. Fig. \ref{fig3}(a) results were obtained for $\nu=7.8$ GHz. Solid curve corresponds to the incidence angle equal to the Brewster angle at such frequency value ($\theta = \theta_{B} = 33.82^{\circ}$). In contrast, dashed and dotted lines correspond to the incidence angles $\theta = 53.82^{\circ}$ and $\theta = 13.82^{\circ}$, respectively. It is apparent from Fig. \ref{fig3}(a) that the transmission coefficient reaches its maximum value ($T=1$) for discrete values of the normalized input intensity and just when the wave frequency and the incidence angle fulfill the Brewster condition given by Eq. \eqref{eq5}. Such states correspond to stationary solitons within the nonlinear layer, as discussed in previous works \cite{CavalcantiOL2014}. If the angle of incidence is different from the Brewster angle at the given frequency, then the transmission coefficient is always less than one. Consequently, its local maxima do not correspond to soliton states. The frequency dependence of the transmission coefficient depicted in Fig. \ref{fig3}(b) was obtained for the incidence angle $\theta = \theta_B$. Solid, dashed, and dotted curves were computed for the normalized input intensities $a=\alpha \vert E_{i} \vert^2$  = 0.00, 6.01, and 18.12, respectively, corresponding to the maximum transmission at $\nu = 7.8$ GHz and the Brewster angle (cf. solid curve in Fig. \ref{fig3}(a)). Notice that the solid curve, obtained for $a=0$, corresponds to the linear regime. The vertical dashed line in Fig. \ref{fig3}(b) corresponds to the incident-wave frequency at which the Brewster phenomenon occurs. Once again, the maximum-transmission peaks obtained for $a \neq 0$ correspond to soliton-state excitations in the nonlinear layer.

We show in Fig. \ref{fig4} the position dependence of the electric-field amplitude within the optical system obtained for $\nu = 7.8$ GHz. Figs. \ref{fig4}(a), \ref{fig4}(b), and \ref{fig4}(c) correspond to the electromagnetic states associated with the maximum values of the transmission coefficient displayed in Fig. \ref{fig3}(a) for $\theta=13.82^{\circ}<\theta_B$,  $\theta=33.82^{\circ}=\theta_B$, and $\theta=53.82^{\circ}>\theta_B$, respectively. The white and shadow areas in Fig. \ref{fig4} correspond to the linear and nonlinear layers, respectively. From Fig. \ref{fig4}(b), one may see the existence of stationary solitons within the nonlinear layer, whereas the electric-field amplitude remains constant within the linear material. Such behavior is consistent with the Brewster effect at the interface of incidence. The present results suggest that the Brewster phenomenon leads to the formation of solitons within the system for oblique incidence. 

\subsection{Single-layer nonlinear medium}

\begin{figure}
\centering
\begin{tabular}{rcl}
\includegraphics[width=0.28\linewidth]{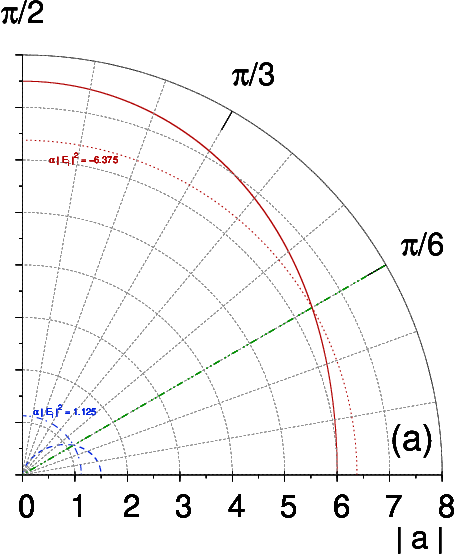} &
\includegraphics[width=0.33\linewidth]{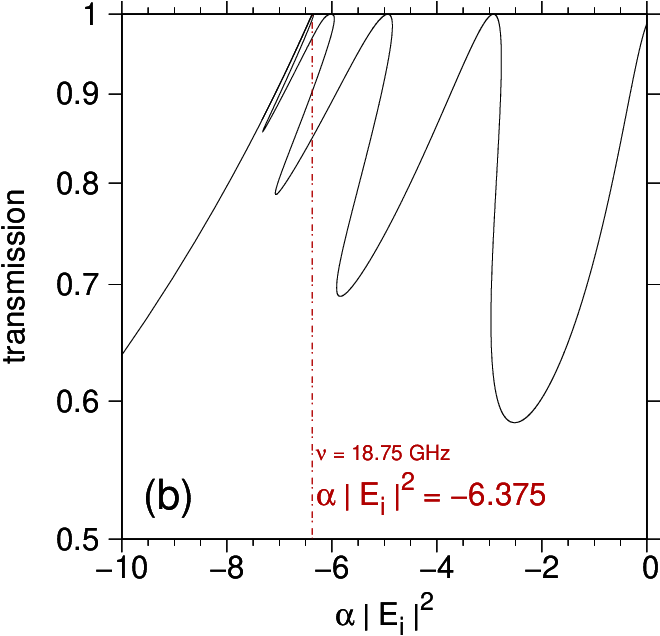} &
\includegraphics[width=0.33\linewidth]{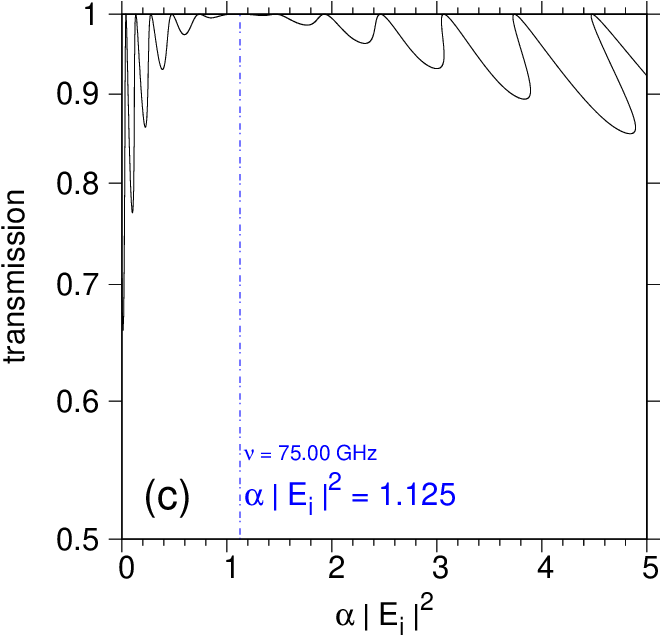} 
\end{tabular}
\caption{(Color online) (a) The Brewster lines $\theta_B = \theta_B (a)$. Results are displayed as functions of $\vert a \vert$. Solid and dashed lines correspond to $\varepsilon_{nl}^0 = 8.0$ and $\varepsilon_{nl}^0 = 0.5$, respectively. Doted and doted-dashed lines correspond to the values of the normalized input intensity corresponding to the Brewster angle $\theta_B = \pi/6$ (see the oblique dotted-dotted-dashed line in panel (a)). Results of panels (b) and (c) correspond to the transmission coefficients, as functions of the normalized input intensity,  for $\varepsilon_{nl}^0 = 8.0$ and $\varepsilon_{nl}^0 = 0.5$, respectively, and the incidence angle $\theta = \pi/6$. Vertical dotted-dashed lines in panels (b) and (c) correspond to the values of the normalized input intensity at which the Brewster effect occurs (see the dotted and dotted-dashed lines in panel (a)). We take the nonlinear layer of thickness $d_{nl} = 10.0$ mm and the magnetic permeability  $\mu_{nl} = 2.0$ in all panels. The wave frequency is computed from the expression $\lambda = 0.4 d_{nl}$, where $\lambda \sqrt{ \vert \varepsilon_{nl}^0 \mu_{nl} \vert}  = c / \nu$ \cite{ChenPRB1987-1}.}
\label{fig6}
\end{figure}

\begin{figure}
\centering
\begin{tabular}{rl}
\includegraphics[width=0.5\linewidth]{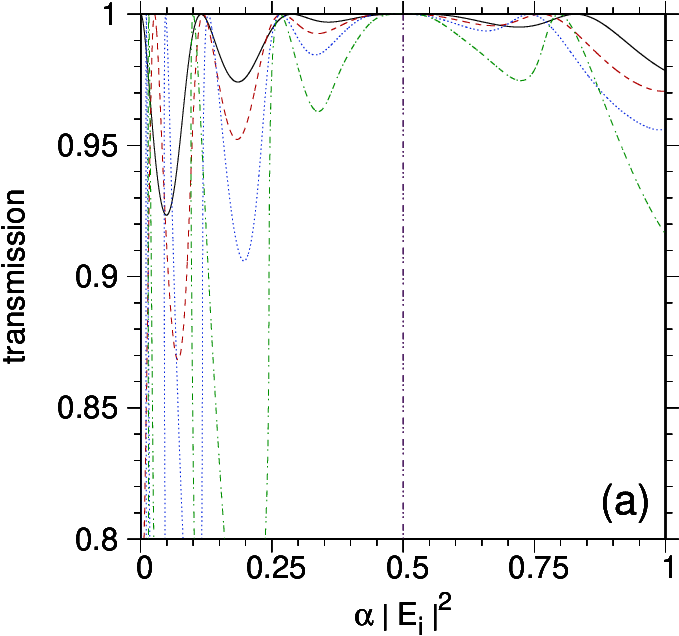} &
\includegraphics[width=0.5\linewidth]{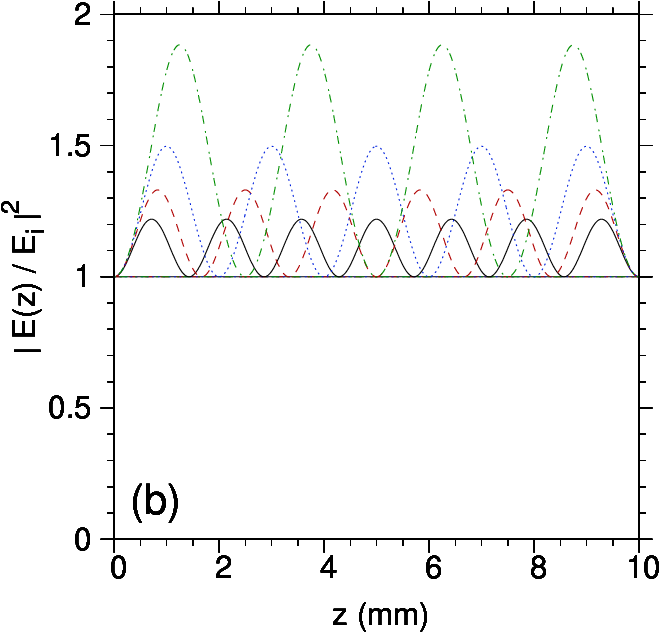}
\end{tabular} 
\caption{(Color online) (a) Transmission coefficients, as functions of the normalized input intensity,  for $\mu_{nl} = \mu_0 = 1$ and $\varepsilon_{nl}^0 = 0.5$. The solid line corresponds to the case of normal incidence, whereas dashed, dotted, and dotted-dashed lines correspond to the incidence angles $\theta = \pi/6$, $\theta = \pi/4$, and $\theta = \pi/3$, respectively. The vertical dotted-dashed line corresponds to the value of the normalized input intensity at which $T=1$. (b) The electric-field amplitude of the electromagnetic wave within the nonlinear layer. Solid, dashed, dotted, and dotted-dashed curves above the horizontal line correspond to $\theta=0$, $\theta = \pi/6$, $\theta = \pi/4$, and $\theta = \pi/3$, respectively, and were computed for the corresponding transmission peaks at $a=0.2884$, $a=0.2690$, $a=0.2674$, and $a=0.2654$, respectively (cf. panel (a)). The horizontal line corresponds to the superposition of all the results obtained for $\theta=0$, $\theta = \pi/6$, $\theta = \pi/4$, and $\theta = \pi/3$, but for $a=0.5$ where the omnidirectional soliton occurs. We have taken $d_{nl} = 10.0$ mm and calculated the wave frequency from the condition $\lambda = 0.4 d_{nl}$, where $\lambda \sqrt{ \vert \varepsilon_{nl}^0 \mu_{nl} \vert}  = c / \nu$ \cite{ChenPRB1987-1}, resulting in $\nu = 106.066$ GHz.}
\label{fig7}
\end{figure}

Now, we consider the optical system as depicted in Fig. \ref{fig1}(b). In this case, the Eq. \eqref{eq10} governs the Brewster phenomenon. We show in Fig. \ref{fig5} the background electric permittivity $\varepsilon_{nl}^0$ as a function of the magnetic permeability $\mu_{nl}$ to fulfill the Brewster condition given by Eq. \eqref{eq10}. Results were obtained for two different values of the incidence angle. Solid curves correspond to the contour lines of the normalized input intensity $a = \alpha \vert E_i \vert^2$. It is apparent from Fig. \ref{fig5} that the Brewster phenomenon depends on the relation between $\varepsilon_{nl} (z_1)/\varepsilon_0$ and  $\mu_{nl} (z_1)/\mu_0$, and may occur in self-focusing and self-defocusing optical materials. We emphasize that Eq. \eqref{eq10} implicitly defines the Brewster angle as a function of $\varepsilon_{nl}^0$, $\mu_{nl}$, $\varepsilon_0$, $\mu_0$, and $a$. If all the optical parameters describing the nonlinear material depend on the wave frequency, then the Brewster angle is a function of $\nu$ and $a$. One has $\theta_B = \theta_B (a)$ for non-dispersive materials.

To exemplify the occurrence of the Brewster phenomenon in the case of nonlinear layers, we have considered two different hypothetical materials. We have supposed the optical constants to be independent of the wave frequency. For the first material we have set $\varepsilon_{nl}^0 = 8.0$, whereas for the second material $\varepsilon_{nl}^0 = 0.5$. In all cases we have considered $\mu_{nl} = 2.0$, $d_{nl} = 10.0$ mm. The wave frequency was obtained from the expression  $\lambda = 0.4 d_{nl}$, where $\lambda \sqrt{\vert \varepsilon_{nl}^0 \mu_{nl} \vert}  = c / \nu$ \cite{ChenPRB1987-1}, and $c$ is the speed of light in vacuum ($\nu=18.75$ GHz and $\nu=75.00$ GHz for the first and second material layers, respectively). We display in Fig. \ref{fig6}(a) the Brewster line $\theta_B = \theta_B (a)$. The results are depicted as functions of $\vert a \vert$ to describe both the self-focusing and self-defocusing cases in the same graph. As described above, solid and dashed lines correspond to layers of the first and second optical materials. Of course, the condition $\vert \sin (\theta) \vert^2 \leq 1 $ in Eq. \eqref{eq10} limits the range of $a$ for which the Brewster effect occurs. For the optical constants of the first material, the Brewster effect takes place for $-7.5 \leq a \leq - 6$, i.e., for a self-defocusing Kerr nonlinearity. For the second set of optical constants, the Brewster effect occurs for self-focusing nonlinear materials such that $0 \leq a \leq 1.5$.

We show in Fig. \ref{fig6}(b) and \ref{fig6}(c) the transmission coefficients as functions of the normalized input intensity, corresponding to layers made of the first and second optical materials, respectively, at the respective values of the wave frequency as described above. Calculations were performed for the incidence angle $\theta = \pi/6$. One may note that for the values of $a$ guarantying the fulfillment of Eq. \eqref{eq10} for $\theta = \pi/6$ (cf. dotted and dotted-dashed lines in Fig. \ref{fig6}(a) and vertical dotted lines in Figs. \ref{fig6}(a) and \ref{fig6}(b)), the transmission coefficients in both Figs. \ref{fig6}(a) and \ref{fig6}(b) exhibit respective peaks, which may be associated with soliton states originating from the Brewster effect. 
 
\subsubsection{Omnidirectional solitons}

Finally, we analyze the case $\mu_{nl} = \mu_0$. In this scenario, the condition of maximum transmission given by Eq. \eqref{eq9} becomes
\be
\label{eq13}
\varepsilon_0 - \varepsilon_{nl}^0 = a.
\ee
The above equation is valid for all possible values of the incidence angle. For the value of the normalized input intensity given by Eq. \eqref{eq13}, an omnidirectional soliton could be obtained. We have numerically verified the above statement by choosing $\mu_{nl} = \mu_0 = 1$, $\varepsilon_0 = 1$, and $\varepsilon_{nl}^0 = 0.5$. Calculations were performed for $d_{nl} = 10$ mm and the wave frequency such that $\lambda = 0.4 d_{nl}$, where $\lambda \sqrt{ \vert \varepsilon_{nl}^0 \mu_{nl} \vert}  = c / \nu$ ($\nu = 106.066$ GHz in the present case). Numerical results for the transmission coefficients obtained for various values of the incidence angle are depicted in Fig. \ref{fig7}(a). The existence of an omnidirectional soliton states around $a=0.5$, the value of $a$ satisfying Eq. \eqref{eq13}, is apparent from Fig. \ref{fig7}(a).

The physical interpretation of this result is as follows: the normalized input intensity value $a$  in Eq. \eqref{eq13} compensates the dielectric constant of the nonlinear medium to match the value of $\varepsilon_0$ of the surrounding medium. Concerning the wave propagation, the nonlinear medium behaves precisely like the surrounding medium for this specific value of $a$, i.e., it becomes homogeneous to the propagation of the electromagnetic wave, and it is impossible to distinguish the surrounding medium from the nonlinear layer. Thus, the formation of the soliton state can be understood in terms of the homogenization of the optical properties of the system induced by the intensity of the incident wave. As the system operates under the condition $\mu_{nl} = \mu_0$, the above-described phenomenon is independent of the incidence angle. Consequently, the amplitude of the electromagnetic field of the wave in the nonlinear layer should be independent of position, as one may note from Fig. \ref{fig7}(b). Just for comparison purposes, we have also shown in Fig. \ref{fig7}(b) the bright-soliton waves excited at the transmission peaks at $a=0.2884$, $a=0.2690$, $a=0.2674$, and $a=0.2654$, corresponding to $\theta=0$, $\theta = \pi/6$, $\theta = \pi/4$, and $\theta = \pi/3$, respectively. Such states do not match Eq. \eqref{eq13} and, therefore, do not correspond to omnidirectional solitons.

\section{Conclusions}
\label{conclusions}

In summary, we have performed a theoretical study on the possibility of exciting stationary solitons due to the Brewster effect occurring in linear/nonlinear optical systems and single nonlinear layers. We have considered Kerr nonlinearity in both systems. In the case of the linear/nonlinear system, we have supposed that the TE-polarized electromagnetic wave is obliquely incident from the vacuum on the linear layer. By combining the transfer-matrix formalism and the numerical solution of the Maxwell equations within the nonlinear layer, we computed the transmission coefficient and analyzed the occurrence of maximum-transmission states. The obtained numerical results suggest the excitation of stationary solitons in the nonlinear layer induced by the occurrence of the Brewster phenomenon in the linear medium.

For the optical system consisting of single nonlinear layers, we exploited the analytical solution outlined in Ref. \onlinecite{HincapieEPL2021} to extend the Brewster phenomenon to include the effects of the normalized input intensity. In the case of non-dispersive nonlinear materials, we have demonstrated that the Brewster phenomenon occurs just for specific values of the normalized input intensity, fulfilling the Brewster condition, and the Brewster angle depends on the normalized input intensity. We also showed that matching the magnetic permeability of the surrounding medium and the nonlinear layer leads to the excitation of an omnidirectional soliton. Such excitation occurs for the normalized input intensity that causes the optical homogenization of the electric permittivity throughout the system. These specific results could be of importance for communication technologies.
 
We hope the present paper will be helpful to the scientific community, further experimental studies, and technological applications.

\textbf{Funding:} The present work was partially financed by the Scientific Colombian Agency CODI - University of Antioquia. J. D. Mazo–V\'asquez is part of the Max Planck School of Photonics, supported by the German Federal Ministry of Education and Research (BMBF), the Max Planck Society, and the Fraunhofer Society.


\end{document}